\documentclass[prc,a4paper,showpacs,byrevtex]{revtex4}
\usepackage{graphicx}
\usepackage{dcolumn}
\usepackage{amsmath}
\usepackage{array}
\usepackage{bm}
\usepackage{amssymb}
\usepackage{amsfonts}

\begin{document}

\title{Duality relations in the auxiliary field method}

\author{Bernard Silvestre-Brac}
\email[E-mail: ]{silvestre@lpsc.in2p3.fr}
\affiliation{LPSC Universit\'{e} Joseph Fourier, Grenoble 1,
CNRS/IN2P3, Institut Polytechnique de Grenoble, 
Avenue des Martyrs 53, F-38026 Grenoble-Cedex, France}
\author{Claude Semay}
\email[E-mail: ]{claude.semay@umons.ac.be}
\affiliation{Service de Physique Nucl\'{e}aire et Subnucl\'{e}aire,
Universit\'{e} de Mons - UMONS, 
Acad\'{e}mie universitaire Wallonie-Bruxelles, 
Place du Parc 20, 7000 Mons, Belgium }

\date{\today}

\begin{abstract}
The eigenenergies $\epsilon^{(N)}(m;\{n_i,l_i\})$ of a system of $N$ identical particles
with a mass $m$ are functions of the various radial quantum numbers $n_i$ and orbital
quantum numbers $l_i$. Approximations $E^{(N)}(m;Q)$ of these eigenenergies, depending
on a principal quantum number $Q(\{n_i,l_i\})$, can be obtained in the framework of the
auxiliary field method. We demonstrate the existence of numerous exact duality relations
linking quantities $E^{(N)}(m;Q)$ and $E^{(p)}(m';Q')$ for various forms of the potentials
(independent of $m$ and $N$) and for both nonrelativistic and semirelativistic kinematics.
As the approximations computed with the auxiliary field method  can be very close to the
exact results, we show with several examples that these duality relations still hold, with
sometimes a good accuracy, for the exact eigenenergies $\epsilon^{(N)}(m;\{n_i,l_i\})$.
\end{abstract}

\maketitle

\section{Introduction}
\label{sec:intro}
The auxiliary field method (AFM) is a very powerful method to obtain analytical
expressions for the eigenvalues of one, two and many-body systems. At the beginning,
it was introduced to deal, in a simple way, with the worrying square root operator
appearing in many relativistic theories \cite{dir66,brink77,deser76,polya81}. But
we showed, in a series of papers \cite{bsb08a,bsb08b,AFMeigen,bsb09a,Sem09a,bsb09c,silv10},
that it can be applied as well with great success in many physical situations. 
In this work, we focus mainly on the properties
of the eigenenergies of $N$-body systems and the relations linking them. So,
we will recall only the basic ingredients necessary for the understanding of the subject
treated here, and we refer the reader to our review paper \cite{bsb10b} for an exhaustive
overview of the method and its applications. Many detailed
discussions on the AFM properties for $N$-body systems can also be found in \cite{silv10}.

A great advantage of the AFM lies in its great simplicity and its ability to treat on
equal footing the ground state and the various excited states. An analytical expression for 
these eigenenergies (the mass $M$ for a semirelativistic approach or the binding energies $E$
for a nonrelativistic one) is particularly interesting from
two points of view:
\begin{itemize}
	\item It allows a very fast calculation in terms of the parameters and thus
	is very well suited if such calculations are a part of a chi-square
	determination of these parameters, a procedure which needs a lot of calls to the
	calculation of the energies.
	\item It provides the behavior of the eigenenergies as functions of the various
	parameters and quantum numbers.
\end{itemize}
This last item can be invoked to search for relationships between the eigenenergies
of different systems with different sets of parameters: we call these relationships
``duality relations''. They form the main subject of this paper.

In essence, the AFM replaces a problem which is not solvable, for example because of
a complicated potential $V(r)$, by another one which can be treated analytically, for
example by the use of a more convenient potential $P(r)$. In so doing, it is necessary
to introduce an auxiliary field $\hat{\nu}$ which is an operator. The original
Hamiltonian $H$ is replaced by a new simpler Hamiltonian $\tilde{H}(\nu)$, called the
AFM Hamiltonian. If this auxiliary field is chosen as $\hat{\nu}_0$ in order to extremize the AFM
Hamiltonian, it can be shown that the AFM Hamiltonian coincides with the original
Hamiltonian: $\tilde{H}(\hat{\nu}_0)=H$. Thus, both formulations are completely equivalent.

The approximation lies in the fact that the auxiliary field is considered no longer as an
operator, but as a real constant $\nu$. The value $\nu_0$ which extremizes the eigenenergy
$E(\nu)$ of the AFM Hamiltonian is then inserted in the corresponding expression to give
an approximate value $E(\nu_0)$ of the exact eigenenergy $E$. An approximate state
for the corresponding eigenstate can also be obtained. The quality of this approximation
has been studied and discussed in detail in the papers mentioned above.

In this paper, we are concerned with systems composed of $N$ particles interacting via
one-body potentials $U_i$ and two-body potentials $V_{ij}$, and moving with a nonrelativistic or a 
semirelativistic kinetic energy. Although the AFM can be invoked to treat this problem in such a general form,
it is manageable in practice only if the particles are identical. This implies that they all
have the same mass $m$, that the form of the one-body potentials is the same for all particles
$U_i \equiv U$, and that the form of the two-body potentials is the same for all pairs of
particles $V_{ij} \equiv V$. In consequence, the Hamiltonian under consideration in this work
has the following form
\begin{equation}
\label{eq:orhamil}
H = \sum_{i=1}^N \sqrt{\bm p_i^2 + m^2} + \sum_{i=1}^N U(|\bm r_i - \bm R|) +
    \sum_{i<j = 1}^N V(|\bm r_i - \bm r_j |).
\end{equation}
In this expression $\bm r_i$ is the position operator for particle $i$, $\bm p_i$ its
conjugate momentum and $\bm R$ the position of the center of mass of the $N$ particles.

The only case which is entirely solvable analytically is a nonrelativistic system with quadratic
potentials $U(r) = V(r)= r^2$. With the natural choice $P(r) = r^2$, 
the AFM needs two auxiliary functions (the prime symbol
stands for the derivative with respect to the argument) $K(r)$ = $U'(r)/P'(r)$ = $U'(r)/(2r)$
and $L(r)$ = $V'(r)/P'(r)$ = $V'(r)/(2r)$, and the introduction of three auxiliary fields:
$\mu$ to replace the relativistic kinetic energy by a nonrelativistic one, $\nu$ to replace
the potential $U(r)$ by the potential $P(r)=r^2$ and $\rho$ to replace the potential $V(r)$
by the potential $P(r)=r^2$. The minimization procedure giving the best values $\mu_0$, $\nu_0$,
$\rho_0$ leads to a system of 3 coupled non linear equations (see \cite{silv10}).

Introducing the new variable

\begin{equation}
\label{eq:defX0}
X_0 = \sqrt{2 \mu_0 (\nu_0 + N \rho_0)},
\end{equation}
it appears that the previous system can be recast under the form of a \textbf{single
transcendental equation} which looks like
\begin{equation}
\label{eq:trans1}
X_0^2 = 2 \sqrt{m^2+\frac{Q}{N}X_0} \left[K \left( \sqrt{\frac{Q}{NX_0}} \right)
        + N L \left( \sqrt{\frac{2Q}{(N-1)X_0}} \right) \right],
\end{equation}
while the corresponding AFM eigenmass is given by
\begin{equation}
\label{eq:genmass}
M(X_0)=N \sqrt{m^2+\frac{Q}{N}X_0} + N U \left( \sqrt{\frac{Q}{NX_0}} \right) + C_N
       V \left( \sqrt{\frac{2Q}{(N-1)X_0}} \right),
\end{equation}
where the number of pairs
\begin{equation}
\label{eq:defcn}
C_N = \frac{N(N-1)}{2}
\end{equation}
has been introduced for convenience.

In these expressions, $Q$ denotes the principal quantum number. Concerning this point,
some comments are in order. The system being described by $(N-1)$ Jacobi internal
coordinates, the most general excited state depends on $(N-1)$ radial quantum numbers
$n_i$ and $(N-1)$ orbital quantum numbers $l_i$, as well as intermediate coupling
quantum numbers which are not considered here. The AFM relying on a quadratic potential,
the principal quantum number resulting
from the AFM treatment is naturally
\begin{equation}
\label{eq:princnumoh}
Q = \sum_{i=1}^{N-1} (2 n_i + l_i) + \frac{3}{2} (N - 1).
\end{equation}
In particular, the ground state for a $N$-boson system is just $Q = 3(N-1)/2$, while
the ground state of a $N$-fermion system is much more involved and needs the introduction
of the Fermi level \cite{silv10}.

In a two-body system, $Q$ reduces to $2 n+l+3/2$. But it is possible in this case to
use other forms for $P(r)$ leading to other expressions for $Q$. For instance, $Q=n+l+1$
with $P(r)=-1/r$ \cite{bsb08a} and $Q=2 ( -\alpha_n/3 )^{3/2}$ for S-states with
$P(r)=r$, where $\alpha_n$ is the $(n+1)^{\textrm{th}}$ zero of the Airy function Ai \cite{bsb10b}.
Moreover, it has been shown (for $N=2$ in \cite{bsb08a} and for arbitrary $N$ in \cite{silv10}) 
that a much better approximation of the exact
energies can be obtained with a slight modification of the principal quantum number. A
particularly simple form which works quite well is given by
\begin{equation}
\label{eq:princnumod}
Q = \sum_{i=1}^{N-1} (\alpha_i n_i + \beta _i l_i) + \gamma.
\end{equation}
In the following, we consider the principal quantum number $Q$ as a variable by its own in
all the computations, and only at the very end of the calculation we replace this undetermined
variable by the most precise one $Q = Q(\{n_i,l_i\})$ as in (\ref{eq:princnumoh}) or in
(\ref{eq:princnumod}).

It is remarkable that, for such a complicated and sophisticated system, the only difficulty
to get the AFM approximate solution needs to solve the transcendental equation (\ref{eq:trans1}),
a quite easy procedure numerically. Moreover, in a number of interesting cases, this solution is
analytical so that the mass itself (given by (\ref{eq:genmass})) is also analytical.

In general, the potential $U(r)$ (or $V(r)$) depends on physical parameters $\{ \tau_1,\tau_2,
\ldots, \tau_p \} \equiv \{ \tau \}$ and should be noted more precisely $U(\{ \tau \};r)$. The
other parameters of the problem are the number of particles $N$ and the mass $m$ of the particles. 
Lastly, we want to have a description of the whole spectrum, so that the principal
quantum number $Q$ also enters the game. In consequence, a complete notation for the eigenmasses
would be $M^{(N)}(\{ \tau \};m,Q)$.

Let us stress now a very important point: for realistic potentials, the parameters $\{ \tau \}$
could depend on $N$ or/and $m$. We do not consider such a particular behavior here. Thus, in
this paper, we assume that \textbf{the parameters of the potentials are independent of $N$ and
$m$}. This means that the $m$ dependence of $H$ is only through the kinetic energy and its $N$
dependence through the numbers of terms in the summation. In this paper, we suppose that the
potentials are given once for all and consider that their form and parameters do not vary for
all studied systems. Therefore we are finally interested only by the $N$, $m$, $Q$ dependence
of the eigenmasses and use the simplified notation $M^{(N)}(m,Q)$.

The two-body problem, $N=2$, deserves a special treatment. In this case $\bm R = (\bm r_1 +
\bm r_2)/2$ so that $|\bm r_1 - \bm R|$ = $|\bm r_2 - \bm R|$ = $(1/2)|\bm r_1 - \bm r_2|$. 
Thus, $U(|\bm r_1 - \bm R|) + U(|\bm r_1 - \bm R|) = 2 U(r/2)$ where $r$ is the inter-distance 
between the particles. We can then write $2 U(r/2) = V(r)$. In
consequence, the distinction between one-body and two-body potentials becomes irrelevant and
it is sufficient to deal with only one type of potential and, depending on the situation under
consideration, we will choose the most convenient one for our purpose. 

Moreover, the semirelativistic
kinetic energy writes ($\bm p$ is the conjugate momentum of the relative distance $\bm r$)
$T = \sigma \sqrt{\bm p^2 + m^2}$, with $\sigma = 1$ for one-body systems and $\sigma = 2$ for
two-body systems. However, we found very convenient to consider the $\sigma$ parameter as a
free real parameter, because the eigenenergy can be calculated as well as a function of $\sigma$.
This possibility leads to interesting properties. Thus, eigenmasses of the systems with the
previous form of the kinetic energy will be noted $M^{(s)}(\sigma,m,Q)$ while the notation
$M^{(2)}(m,Q)$ is reserved to the natural definition $M^{(N=2)}(m,Q)$ = $M^{(s)}(2,m,Q)$.

What we call a ``duality relation'' is just a relation between $M^{(N)}(m,Q)$, $M^{(p)}(m',Q')$
and $M^{(s)}(\sigma,m'',Q'')$. The search for such duality relations is the subject of this paper.
Of course, these duality relations are exact only for the AFM approximation of a given problem.
However, since this approximation may be, under many circumstances, a rather accurate result
of the exact eigenvalues, our hope is that these duality relations remain approximately valid
for the exact results and thus represent a rough view of reality. In particular, starting
with the expression for a two-body system, whose calculation is rather easy, one can hope that the
main trends for a $N$-body system are reachable as the consequence of the duality relations. 

In the next section, we discuss duality relations in the general case, while in the 3rd and
4th sections, we investigate the particular cases of ultrarelativistic systems and
nonrelativistic systems, respectively. Their interconnexion is investigated in detail in the 5th section.
In the 6th section, applications are given while conclusions are drawn in the last section.  
Some checks of the duality relations obtained are performed with explicit AFM solutions in 
the Appendix.

\section{General Case}
\label{sec:gencase}

\subsection{Compact form}
\label{sec:compf}
In this section, we show that it is possible to solve formally the transcendental equation
(\ref{eq:trans1}) and express the AFM mass (\ref{eq:genmass}) under a more compact form.
In order to do that, we switch from the $X_0$ variable to the $y$ variable defined by
\begin{equation}
\label{eq:defy}
y=m \sqrt{\frac{N}{QX_0}}.
\end{equation}
With this definition, it is easy to calculate the following quantities:
\begin{equation}
\label{eq:defquants}
\sqrt{m^2+\frac{Q}{N}X_0} =  \frac{m}{y} \sqrt{1+y^2}, \quad 
\sqrt{\frac{Q}{NX_0}} =  \frac{Q}{mN} y, \quad
\sqrt{\frac{2Q}{(N-1)X_0}} =  \frac{Q}{m \sqrt{C_N}} y.
\end{equation}
Let us introduce now, instead of potentials $U$ and $V$, the related potentials $W$, $Y$ and
$Z$ by
\begin{eqnarray}
\label{eq:defnewpot}
U \left( \frac{Q}{mN} x \right)& = & C_N W(x), \nonumber \\
V \left( \frac{Q}{m \sqrt{C_N}} x \right)& = & N Y(x), \\
Z(x) & = & W(x) + Y(x) \nonumber.
\end{eqnarray}
It is easy to prove that the transcendental equation (\ref{eq:trans1}) can be recast under the
reduced form
\begin{equation}
\label{eq:transred}
\frac{m}{C_N}=y^2 \sqrt{1+y^2} Z'(y).
\end{equation}

The new step of our procedure is the introduction of an inverse function $A$, which
is defined through the relation
\begin{equation}
\label{eq:defA}
A(x^2 \sqrt{1+x^2} Z'(x)) = x.
\end{equation}
One can thus express formally the value of $y$ in term of the $A$ function as
\begin{equation}
\label{eq:yofA}
y = A \left(\frac{m}{C_N} \right).
\end{equation}
The eigenmass follows from (\ref{eq:genmass}) introducing rather the $y$ variable.
Explicitly,
\begin{equation}
\label{eq:massy}
M = N\, m \frac{\sqrt{1+y^2}}{y}  + N\, C_N\, Z(y).
\end{equation}

The last step of our treatment is the introduction of the $B$ function as
\begin{equation}
\label{eq:defB}
B(x) = \frac{x}{A(x)} \sqrt{1+A(x)^2} + Z(A(x)).
\end{equation}
One sees that the eigenmass is expressed simply in a very compact form
\begin{equation}
\label{eq:massB}
M^{(N)}(m,Q) = N C_N B\left(\frac{m}{C_N} \right).
\end{equation}
It is astonishing that the eigenmass of such a complicated system, even in its AFM
approximation, can be expressed in such a compact form in term of a single function
$B(x)$. However, this enthusiastic conclusion must be moderated by the following
remarks:
\begin{itemize}
	\item The $B$ function is expressed through the $A$ function which, itself, needs
	the inversion of a function which contains a square root and a potential with a
	complicated form; this can be a tremendous task. The effort of
	solving the transcendental equation (\ref{eq:trans1}) is replaced by the inversion
	procedure (\ref{eq:defA}), which is formally as much as complicated.
	\item As seen from (\ref{eq:defnewpot}), the $A$ function, and hence the $B$ function,
	depends not only on the considered potentials -- an unavoidable ingredient -- but
	also on $N$, $m$ and $Q$, and should be noted more precisely $A(N,m,Q;x)$. This means that
	this function depends not only on the considered systems (through the potentials, and
	$N$ and $m$ variables) but also on the particular excitation state $Q$.
\end{itemize}

In consequence of these arguments, it appears that this formulation is of no practical
use and do not lead to duality relations (at least we did not succeed to find one!).
However, we gave this demonstration because, independently of the formal esthetical
presentation, it represents the prototype of many similar approaches that will be
extensively used in the following sections.

\subsection{One-body interaction only}
\label{sec:no2bod}
In absence of two-body interaction, $V = 0$, simplifications occur and one can draw
interesting conclusions.
Instead of the $y$ variable, let us use rather the $r_0$ variable (which is called the
mean radius) defined by
\begin{equation}
\label{eq:defr01}
r_0 = \frac{Q}{mN}y =\sqrt{\frac{Q}{N X_0}}.
\end{equation}
With this new variable, the transcendental equation (\ref{eq:transred}) and the mass expression
(\ref{eq:massy}) are replaced by the following ones
\begin{eqnarray}
\label{eq:trans2N}
\frac{Q}{N} & = & r_0^2 \sqrt{1+N^2 (mr_0/Q)^2}\;U'(r_0), \\
\label{eq:mass2N}
M^{(N)}(m,Q) & = & N \left[ \frac{Q}{Nr_0}\sqrt{1+N^2 (mr_0/Q)^2} + U(r_0) \right].
\end{eqnarray}
In this case, we do not want to get a compact closed formula (one thus has to derive $r_0$ from
the first equation giving $r_0(N,m,Q)$ and inserting it in the second equation to get $M^{(N)}(m,Q)$).
We prefer to express in the same form the case of a system of $p$ particles of mass $m'$ interacting
with the \textbf{same one-body potential} $U(r)$ (don't forget that it is $N$ and $m$
independent) in a state of excitation $Q'$. We thus have
\begin{eqnarray}
\label{eq:trans2p}
\frac{Q'}{p} & = & s_0^2 \sqrt{1+p^2 (m's_0/Q')^2}\;U'(s_0), \\
\label{eq:mass2p}
M^{(p)}(m',Q') & = & p \left[ \frac{Q'}{ps_0}\sqrt{1+p^2 (m's_0/Q')^2} + U(s_0) \right].
\end{eqnarray}

Now, let us choose the parameters $m',Q'$ as
\begin{equation}
\label{eq:defmQ2}
m' = m;\quad Q' = \frac{p}{N} Q.
\end{equation}
Inserting these values in (\ref{eq:trans2p}) gives exactly equation (\ref{eq:trans2N}) with $s_0$
in place of $r_0$. We suppose that the function $f(a;x) = x^2 \sqrt{1+a^2 x^2} U'(x)$ is monotonic
in $x$ for physical values of $x$ and $a$ (here $a$ = $m N/Q$); this is always the case in practice.
Since we have $f(a;s_0)$ = $f(a;r_0)$, this implies $s_0 = r_0$. Inserting these values of $m'$,
$Q'$ and $s_0$ in (\ref{eq:mass2p}), it appears that $M^{(p)}$ = $(p/N)M^{(N)}$. We are thus
led to the following duality relation between the $N$-body and $p$-body systems governed by the
same one-body potential
\begin{equation}
\label{eq:dualpN1}
M^{(N)}(m,Q) = \frac{N}{p} M^{(p)}\left( m,\frac{p}{N}Q \right).
\end{equation}
It means that the spectrum of a $N$-body system is the same as the spectrum of a $p$-body system
(with the same particle mass and the same form of the one-body potential) provided we look at
different excitation states. Formula (\ref{eq:dualpN1}) tells that the energy is equally divided on all the particles. This is formally exact for arbitrary values of the principal quantum
number $Q$; however if one relies on the value $Q(\{n_i,l_i\})$ as in (\ref{eq:princnumoh}), there is
no certainty that $pQ/N$ appears still under the form (\ref{eq:princnumoh}). Nevertheless, as we
already stressed, we must make the calculation of the $p$-body system in term of an informal $Q$
and not trying to fit the spectrum of the system, then make the replacement $Q \to
NQ/p$ to get formally $M^{(N)}(m,Q)$ and, lastly, use in this expression the value of
$Q = Q(\{n_i,l_i\})$ that fits at best the spectrum. These comments are important; in the following
we must keep them in mind for all duality relations.

Let us consider a 2-body system, which is the most simple system to be studied. To apply the previous
formula, one must take some caution with the definition of the potential since $V(r) = 2 U(r/2)$. Thus,
if $M^{(2)}(m,Q)$ is the eigenenergy resulting from the Hamiltonian $2 \sqrt{\bm p^2 + m^2} + V(r)$
(which has the form of a spinless Salpeter equation with a two-body potential $V$), the calculation
for $M^{(N)}(m,Q)$ must be done with one-body potential $U(x) = V(2x)/2$. This means that the
determination of the mass of the $N$-body system results rather from the equations
\begin{eqnarray}
\label{eq:trans3N}
\frac{4Q}{N} & = & s_0^2 \sqrt{1+N^2 (m s_0/(2Q))^2}\;V'(s_0), \\
\label{eq:mass3N}
M^{(N)}(m,Q) & = & \frac{N}{2} \left[ \frac{4Q}{N s_0}\sqrt{1+N^2 (m s_0/(2Q))^2} + V(s_0) \right].
\end{eqnarray}

It could be interesting to deal with the quantity $M^{(s)}(\sigma,m',Q')$, eigenvalue of the Hamiltonian
$\sigma \sqrt{\bm p^2 + m'^2} + V(r)$. In this case, $M^{(s)}$ is given by the following set of equations
\begin{eqnarray}
\label{eq:transsN}
\sigma Q' & = & r_0^2 \sqrt{1+(m'r_0/Q')^2}\;V'(r_0), \\
\label{eq:masssN}
M^{(s)}(\sigma,m',Q') & = & \frac{\sigma Q'}{r_0} \sqrt{1+(m'r_0/Q')^2} + V(r_0).
\end{eqnarray}
Putting the value $m' = 2m/\sigma$ and $Q'= 4Q/(\sigma N)$ into the previous equations, one sees that
we recover (\ref{eq:trans3N}) provided that $r_0$ = $s_0$ and, then, $M^{(s)}$ = $(2/N)M^{(N)}$.
Therefore, one has the duality relation
\begin{equation}
\label{eq:dual2N}
M^{(N)}(m,Q) = \frac{N}{2} M^{(s)}\left( \sigma, \frac{2}{\sigma}m,\frac{4}{\sigma N}Q \right).
\end{equation}
This relation is particularly interesting because it is valid whatever the value of $\sigma$. In
particular, the special value $\sigma = 2$ gives the link between $M^{(N)}$ and $M^{(2)}$, as in
(\ref{eq:dualpN1}).

Choosing the value $\sigma = 4/N$ leads to the duality relation
\begin{equation}
\label{eq:dual2Ns}
M^{(N)}(m,Q) = \frac{N}{2} M^{(s)}\left( \frac{4}{N}, \frac{N}{2}m, Q \right).
\end{equation}
In this case, there is a one-to-one correspondence of the spectrum at the price of considering
systems with different particle masses.
The conclusion is that, if $M^{(s)}$ or $M^{(2)}$ can be evaluated analytically, the same property
holds for $M^{(N)}$.

\subsection{Two-body interaction only}
\label{sec:no1bod}

We consider now the case of systems governed only by two-body forces, $U = 0$. One can model
our reasoning on the demonstrations given in the previous section.
In this case the natural variable is the mean radius
\begin{equation}
\label{eq:defr02}
r_0 = \frac{Q}{m \sqrt{C_N}}y =\sqrt{\frac{2 Q}{(N-1)X_0}}.
\end{equation}
With this new variable, the transcendental equation (\ref{eq:transred}) and the mass expression
(\ref{eq:massy}) are replaced by the following ones
\begin{eqnarray}
\label{eq:trans4N}
\frac{2Q}{(N-1)\sqrt{C_N}} & = & r_0^2 \sqrt{1+C_N (m r_0/Q)^2}\;V'(r_0), \\
\label{eq:mass4N}
M^{(N)}(m,Q) & = & C_N \left[ \frac{2Q}{(N-1)\sqrt{C_N}} \frac{1}{r_0}\sqrt{1+C_N (mr_0/Q)^2}
                  + V(r_0) \right].
\end{eqnarray}

Exactly in the same way as the proof given before, one arrives at the following duality relation
between the $N$-body and the $p$-body systems with particles interacting via the \textbf{same two-body
interaction}
\begin{equation}
\label{eq:dualpN4}
M^{(N)}(m,Q) = \frac{C_N}{C_p} M^{(p)}\left( \frac{p-1}{N-1}m,
                \frac{p-1}{N-1}\sqrt{\frac{C_p}{C_N}}\;Q \right).
\end{equation}
In this case, the spectrum of the $N$-body system is the same as the $p$-body system (with the same
two-body potential) provided we consider different particle masses in both situations and for other
excitation states.

The link between $M^{(N)}$ and $M^{(s)}$ now looks like
\begin{equation}
\label{eq:dual2N4}
M^{(N)}(m,Q) = C_N M^{(s)}\left( \sigma, \frac{2}{\sigma (N-1)}m,\frac{2}{\sigma (N-1) \sqrt{C_N}}Q \right).
\end{equation}
Here again, this relation is valid whatever the value of $\sigma$.
In particular, the special value $\sigma = 2$ gives the link between $M^{(N)}$ and $M^{(2)}$, as
resulting from~(\ref{eq:dualpN4}) with $p=2$.

Alternatively, one can use for instance this freedom on $\sigma$ to choose the same mass in both systems
(this was not possible with the expression $M^{(2)}$); indeed, with the choice $\sigma = 2/(N-1)$, one has
\begin{equation}
\label{eq:dual2N5}
M^{(N)}(m,Q) = C_N M^{(s)}\left( \frac{2}{N-1}, m,\frac{Q}{ \sqrt{C_N}} \right).
\end{equation}
Choosing $\sigma = 2/((N-1)\sqrt{C_N})$ allows to express the duality relation keeping the same
principal number in both systems
\begin{equation}
\label{eq:dual2N6}
M^{(N)}(m,Q) = C_N M^{(s)}\left( \frac{2}{(N-1)\sqrt{C_N}}, \sqrt{C_N} m,Q \right).
\end{equation}

\subsection{Link between one and two-body interaction}
\label{sec:no1bodl}

We consider the case of a \textbf{$N$-body system} with particles of mass $m$ and interacting with
a \textbf{two-body interaction $V(x)$ only}. The corresponding mass for a state of excitation number
$Q$, $M^{(N)}(m,Q)$, is given by the set of equations (\ref{eq:trans4N}) and (\ref{eq:mass4N}).
Now we consider the case of a $N$-\textbf{body system} with particles of mass $m'$ and interacting
with \textbf{one-body interaction of the form} $U(x) = c V(x)$ only ($c$ being a constant).
The comparison for the mass of both systems may present some interest: for example in 3-quark
systems, the case $U(x)=ax$ corresponds to the so-called $Y$ confinement potential with the junction
point placed at the center of mass, while the case $V(x)=ax/2$ (in this case $c=1/2$) corresponds
to the alternative description of confinement named the $\Delta$ potential \cite{qcd2}.
In this case, the corresponding mass for a state of excitation number $Q'$, $\tilde{M}^{(N)}(m',Q')$
is given by the set of equations (\ref{eq:trans2p}) and (\ref{eq:mass2p}) with $p=N$.

If one chooses the mass and principal quantum numbers fulfilling the following relations
\begin{eqnarray}
\label{eq:link1mQ}
\frac{2cQ}{(N-1)\sqrt{C_N}} & = & \frac{Q'}{N}, \\
C_N \frac{m^2}{Q^2} & = & N^2 \frac{m'^2}{Q'^2}, \nonumber
\end{eqnarray}
it is easy to show that (\ref{eq:trans2p}) is transformed into (\ref{eq:trans4N}) while
$\tilde{M}^{(N)}/(cN)$ is transformed into $M^{(N)}/C_N$. Therefore we arrive at the following
interesting duality relation
\begin{equation}
\label{eq:dualMNtMN}
M^{(N)}(m,Q) = \frac{N-1}{2c} \tilde{M}^{(N)}\left( \frac{2c}{N-1}m,\frac{4c\sqrt{C_N}}{(N-1)^2} Q 
\right).
\end{equation}

\section{Ultrarelativistic limit}
\label{sec:ultralim}

The case of ultrarelativistic systems, characterized by a vanishing mass $m=0$, presents some
very specific and interesting features. For instance, the case of systems composed of gluons or/and light quarks
can be well represented in this scheme \cite{barlnc}. The Hamiltonian of the system is then
\begin{equation}
\label{eq:urhamil}
H = \sum_{i=1}^N \sqrt{\bm p_i^2} + \sum_{i=1}^N U(|\bm r_i - \bm R|) +
    \sum_{i<j = 1}^N V(|\bm r_i - \bm r_j |).
\end{equation}

Indeed, the formulation is simpler for this particular
situation. Putting the value $m=0$ in (\ref{eq:trans1}) and (\ref{eq:genmass}), one gets a
new set of equations.
The transcendental equation looks like
\begin{equation}
\label{eq:transur}
\sqrt{\frac{N}{Q}} X_0^{3/2} = 2 \left[K \left( \sqrt{\frac{Q}{NX_0}} \right)
        + N L \left( \sqrt{\frac{2Q}{(N-1)X_0}} \right) \right],
\end{equation}
while the corresponding AFM eigenmass is given by
\begin{equation}
\label{eq:urmass}
M_u(X_0)=\sqrt{N Q X_0} + N U \left( \sqrt{\frac{Q}{NX_0}} \right) + C_N
       V \left( \sqrt{\frac{2Q}{(N-1)X_0}} \right).
\end{equation}
The eigenmass, depending now only on $N$ and $Q$, will be noted $M_u^{(N)}(Q)$ (the index $u$
stands for ``ultrarelativistic'').

\subsection{Presence of one and two-body potentials}
\label{sec:urUV}

One can mimic there the approach followed in section \ref{sec:compf}.
Let us introduce, instead of $X_0$, the new variable $y$
\begin{equation}
\label{eq:defyu}
y=\sqrt{\frac{aQ}{X_0}},
\end{equation}
where $a$ is any real parameter.

Let us define now, instead of potentials $U$ and $V$, the related potentials $W$, $Y$ and
$Z$ by
\begin{eqnarray}
\label{eq:defnewpotu}
U \left( \frac{1}{\sqrt{aN}} \; x\right)& = & C_N W(x), \nonumber \\
V \left( \sqrt{\frac{2}{a(N-1)}} \; x \right)& = & N Y(x), \\
Z(x) & = & W(x) + Y(x). \nonumber
\end{eqnarray}
It is easy to prove that the transcendental equation (\ref{eq:transur}) can be recast under the
reduced form:

\begin{equation}
\label{eq:transredu}
\frac{\sqrt{aN}Q}{N C_N} = y^2 Z'(y).
\end{equation}

As in Sect.~\ref{sec:compf}, we introduce $A_u$, the inverse function of $y^2 Z'(y)$,
\begin{equation}
\label{eq:defAu}
A_u(x^2 Z'(x)) = x.
\end{equation}
One can thus express formally the value of $y$ in term of the function $A_u$ as
\begin{equation}
\label{eq:yofAu}
y = A_u \left( \frac{\sqrt{a N} \; Q}{N C_N} \right).
\end{equation}
The eigenmass follows from (\ref{eq:urmass}) introducing rather the $y$ variable.
Explicitly
\begin{equation}
\label{eq:massyu}
M_u^{(N)}(Q) = N C_N \left[\frac{\sqrt{aN} \; Q}{NC_N y} + Z(y) \right].
\end{equation}

Lastly, one defines the function $B_u$ as
\begin{equation}
\label{eq:defBu}
B_u(x) = \frac{x}{A_u(x)} + Z(A_u(x)).
\end{equation}
The eigenmass (\ref{eq:massyu}) is expressed very simply as
\begin{equation}
\label{eq:massBu}
M_u^{(N)}(Q) = N C_N B_u\left( \frac{\sqrt{aN} \; Q}{N C_N} \right).
\end{equation}

As in Sect.~\ref{sec:compf}, we are able to express the ultrarelativistic
eigenmass in a very compact form through a single function $B_u$. Nevertheless, in contrast
to Sect.~\ref{sec:compf}, we have in this case more sympathetic features:
\begin{itemize}
	\item The inversion procedure~(\ref{eq:defAu}) is much easier than the corresponding
	one (\ref{eq:defA}), because of the disappearance of the square root.
	\item The parameter $a$ is free, so that we can choose it in the most convenient way. It
	appears in the argument of the $B_u$ function, but this function itself, as can be seen
	from the definitions~(\ref{eq:defnewpotu}), depends on $a$, so that the final result
	$M_u$ is $a$-independent.
	\item The potentials $W, Y, Z$, and, thus, the function $B_u$ does depend on $U$, $V$,
        $a$ and $N$, but not on $Q$. Therefore the only dependence in $Q$ of the final result is
         through the argument of the $B_u$ function.
\end{itemize}

As a consequence of the last item, for a given system ($U$, $V$ and $N$ are given), if there
is some degeneracy due to $Q$ (for example using the harmonic oscillator expression
(\ref{eq:princnumoh})), this degeneracy persists in the AFM expression~(\ref{eq:massBu}) of
the spectrum. Moreover, if it is possible to obtain an analytical expression for $B_u(x)$,
the behavior of $M_u^{(N)}(Q)$ as function of $Q$ directly follows from (\ref{eq:massBu}).

\subsection{One-body interaction only}
\label{sec:no2bodu}

In absence of two-body interaction, $V = 0$, further simplifications occur and lead to very
interesting conclusions.
Let us choose the special value $a=1/N$ in the previous approach. One has $\sqrt{aN}=1$ and
$Z(x)=U(x)/C_N$. Introducing the mean radius $r_0 = y$, the transcendental equation
(\ref{eq:transredu}) reduces to
\begin{equation}
\label{eq:transredu1}
\frac{Q}{N } = r_0^2 U'(r_0).
\end{equation}

Let us denote $C(x)$ the inverse function of $x^2 U'(x)$,
\begin{equation}
\label{eq:defC}
C(x^2 U'(x)) = x.
\end{equation}
Therefore, the mean radius is given by
\begin{equation}
\label{eq:rofC}
r_0 = C \left( \frac{Q}{N} \right).
\end{equation}
Lastly, introducing the function $F(x)$ by
\begin{equation}
\label{eq:defF}
F(x) = \frac{x}{C(x)} + U(C(x)),
\end{equation}
the mass of the system is given by
\begin{equation}
\label{eq:massF}
M_u^{(N)}(Q) = N F \left( \frac{Q}{N} \right).
\end{equation}

The great advantage of this formulation with respect to the approach of Sect.~\ref{sec:urUV}
is that the $C$ and $F$ functions are \textbf{universal} in the sense that it depends on the form of
the $U$ potential, but it is independent of the system and of the excitation number
(independent of $N,Q$). 
For two different potentials, The $C$ and $F$ functions, although denoted by the same label, are indeed different.

From expression (\ref{eq:massF}), one deduces immediately the duality relation
\begin{equation}
\label{eq:dualuN1}
M_u^{(N)}(Q) = \frac{N}{p} M_u^{(p)} \left( \frac{p}{N} Q\right),
\end{equation}
which is the special case of (\ref{eq:dualpN1}) with $m=0$.

Let us consider the Hamiltonian $H = \sigma \sqrt{\bm p^2} + V(r)$, whose eigenmass is
noted $M_u^{(s)}(\sigma,Q)$. It is easy to show that
\begin{equation}
\label{eq:massFs}
M_u^{(s)}(\sigma,Q) = F \left( \sigma Q \right),
\end{equation}
with the function $F$ as in (\ref{eq:defF}) and $C(x)$ as in (\ref{eq:defC}) but calculated
with the potential $V(x)$.

The universality of $F(x)$ does not bring fundamental new results. In practice, all the
properties derived in Sect.~\ref{sec:no2bod} can be invoked with the special choice $m=0$ to
get simplified formulations.
For example, duality relation (\ref{eq:dual2N}) writes now
\begin{equation}
\label{eq:dual2Nu}
M_u^{(N)}(Q) = \frac{N}{2} M_u^{(s)}\left( \sigma, \frac{4}{\sigma N}Q \right),
\end{equation}
valid whatever the value of $\sigma$. 
Putting $\sigma = 2$ in this relation, one recovers (\ref{eq:dualuN1}) with $p=2$. Using
$\sigma=4/N$, one obtains the following interesting duality relation
\begin{equation}
\label{eq:dual2Nuu}
M_u^{(N)}(Q) = \frac{N}{2} M_u^{(s)}\left( \frac{4}{N}, Q \right),
\end{equation}
making a one-to-one correspondence in the spectrum of both systems.

\subsection{Two-body interaction only}
\label{sec:no2bodv}

In absence of one-body interaction, $U = 0$, we have also interesting conclusions.
Let us choose the special value $a=2/(N-1)$ in the previous approach. One has
$\sqrt{aN}=N/\sqrt{C_N}$ and $Z(x)=V(x)/N$. Introducing the mean radius $r_0 = y$,
the transcendental equation (\ref{eq:transredu}) reduces to
\begin{equation}
\label{eq:transredu2}
\frac{2Q}{(N-1) \sqrt{C_N}} = r_0^2 V'(r_0).
\end{equation}

Defining the $C(x)$ and $F(x)$ function as in (\ref{eq:defC}) and (\ref{eq:defF})
calculated with the potential $V(x)$, one has
\begin{eqnarray}
\label{eq:rofC2}
r_0 & = & C \left( \frac{2Q}{(N-1) \sqrt{C_N}} \right), \\
\label{eq:massF2}
M_u^{(N)}(Q) & = & C_N F \left( \frac{2Q}{(N-1) \sqrt{C_N}} \right).
\end{eqnarray}
As the previous case studied in Sect.~\ref{sec:no2bodu}, the $C$ and $F$ functions are \textbf{universal}.
Let us note that (\ref{eq:massF2}) with $N=2$ coincides with (\ref{eq:massFs}) for $\sigma=2$, as expected.

From expression (\ref{eq:massF2}), one deduces immediately the duality relation
\begin{equation}
\label{eq:dualuN2}
M_u^{(N)}(Q) = \frac{C_N}{C_p} M_u^{(p)} \left( \frac{p-1}{N-1} \sqrt{\frac{C_p}{C_N}}Q\right),
\end{equation}
which is the special case of (\ref{eq:dualpN4}) with $m=0$.

Application of (\ref{eq:dual2N4}) with $m=0$ provides the duality relation
\begin{equation}
\label{eq:dual2Nv}
M_u^{(N)}(Q) = C_N M_u^{(s)}\left( \sigma, \frac{2}{\sigma (N-1) \sqrt{C_N}}Q \right),
\end{equation}
valid whatever the value of $\sigma$. 
Putting $\sigma = 2$ in this relation, one recovers (\ref{eq:dualuN2}) with $p=2$. Using
$\sigma=2/((N-1)\sqrt{C_N})$, one obtains the following interesting duality relation
\begin{equation}
\label{eq:dual2Nuub} 
M_u^{(N)}(Q) = C_N M_u^{(s)}\left( \frac{2}{(N-1)\sqrt{C_N}}, Q \right),
\end{equation}
making a one-to-one correspondence in the spectrum of both systems.

\subsection{Link between one and two-body interaction}
\label{sec:no1bodu}

With the same conditions of application studied in section \ref{sec:no1bodl}, the interesting relation
(\ref{eq:dualMNtMN}) looks even simpler
\begin{equation}
\label{eq:dualMNtMNu}
M_u^{(N)}(Q) = \frac{N-1}{2c} \tilde{M}_u^{(N)}\left( \frac{4c\sqrt{C_N}}{(N-1)^2} Q \right).
\end{equation}

\section{Nonrelativistic limit}
\label{sec:nrlim}
Another interesting limit of the theory is the nonrelativistic one, valid when the mass of
the particles is large compared to the mean potential. In this case, the considered Hamiltonian
is simply
\begin{equation}
\label{eq:orhamilnr}
H = \sum_{i=1}^N \frac{\bm p_i^2}{2m} + \sum_{i=1}^N U(|\bm r_i - \bm R|) +
    \sum_{i<j = 1}^N V(|\bm r_i - \bm r_j |).
\end{equation}
Instead of dealing with the total mass $M^{(N)}$, it is better to consider the binding energy
obtained by removing the total rest mass: $E^{(N)} = M^{(N)} - Nm$. The AFM approximation
$E^{(N)}(m,Q)$ of the binding energy is given by the following equations,
\begin{equation}
\label{eq:transnr}
X_0^2 = 2 m \left[K \left( \sqrt{\frac{Q}{NX_0}} \right)+ 
        N L \left( \sqrt{\frac{2Q}{(N-1)X_0}} \right) \right]
\end{equation}
and
\begin{equation}
\label{eq:nrmass}
E^{(N)}(X_0)= \frac{QX_0}{2m}+ N U \left( \sqrt{\frac{Q}{NX_0}} \right) +
        C_N V \left( \sqrt{\frac{2Q}{(N-1)X_0}} \right).
\end{equation}
We will see that, in this nonrelativistic limit, one obtains additional interesting properties.

\subsection{Presence of one and two-body potentials}
\label{sec:nrUV}

Let us introduce the same parameter $y$ as in (\ref{eq:defyu}) and the same $Z(x)$ function as
in (\ref{eq:defnewpotu}). The transcendental equation looks like
\begin{equation}
\label{eq:transredn}
\frac{a Q^2}{m N C_N} = y^3 Z'(y).
\end{equation}

We introduce $A_n$, the inverse function of $y^3 Z'(y)$
\begin{equation}
\label{eq:defAn}
A_n(x^3 Z'(x)) = x.
\end{equation}
The value of $y$ in term of the function $A_n$ is expressed as
\begin{equation}
\label{eq:yofAn}
y = A_n \left( \frac{a Q^2}{m N C_N} \right).
\end{equation}

The eigenenergy follows from (\ref{eq:nrmass}) introducing rather the $y$ variable.
Explicitly
\begin{equation}
\label{eq:Eyn}
E^{(N)}(m,Q) = N C_N \left[\frac{a Q^2}{2m NC_N y^2} + Z(y) \right].
\end{equation}
Lastly, one defines the function $B_n$ as
\begin{equation}
\label{eq:defBn}
B_n(x) = \frac{x}{2 A_n(x)^2} + Z(A_n(x)).
\end{equation}
The binding energy (\ref{eq:Eyn}) is expressed very simply as
\begin{equation}
\label{eq:EBn}
E^{(N)}(m,Q) = N C_N B_n\left( \frac{a Q^2}{m N C_N} \right).
\end{equation}

As in the case of ultrarelativistic limit, the function $B_n$ does depend on $U$, $V$,
$a$ and $N$, but not on $m$ and $Q$. Thus the conclusion concerning the degeneracy of the
spectrum remains valid. Moreover, if one knows an analytical expression of $B_n(x)$ then
(\ref{eq:EBn}) provides the behavior of $E^{(N)}(m,Q)$ as a function of $m$ and $Q$.

In the case of a nonrelativistic system, an additional property appears. Remarking
that $E^{(N)}$ depends only on the ratio $Q^2/m$, one has the general duality relation
\begin{equation}
\label{eq:dualgennr}
E^{(N)}(m,Q) = E^{(N)}(\beta^2 m, \beta Q),
\end{equation}
valid for any value of the real parameter $\beta$. This relation is very strong because it
is completely general, even if the system is governed by the presence of both one
\textbf{and} two-body interactions.

\subsection{One-body interaction only}
\label{sec:no2bodnru}

As in Sect.~\ref{sec:no2bodu}, the absence of two-body interaction, $V = 0$,
implies further simplifications. 
Let us choose the special value $a=1/N$ in the previous approach. One has $\sqrt{aN}=1$ and
$Z(x)=U(x)/C_N$. Introducing the mean radius $r_0 = y$, the transcendental equation
(\ref{eq:transredn}) reduces to
\begin{equation}
\label{eq:transredn1}
\frac{Q^2}{m N^2} = r_0^3 U'(r_0).
\end{equation}

Let us denote $D(x)$ the inverse function of $x^3 U'(x)$
\begin{equation}
\label{eq:defD}
D(x^3 U'(x)) = x.
\end{equation}
Therefore, the mean radius is given by
\begin{equation}
\label{eq:rofD}
r_0 = D \left( \frac{Q^2}{m N^2} \right).
\end{equation}
Lastly, introducing the function $G(x)$ by
\begin{equation}
\label{eq:defG}
G(x) = \frac{x}{2 D(x)^2} + U(D(x)),
\end{equation}
the binding energy of the system is given by
\begin{equation}
\label{eq:EF}
E^{(N)}(m,Q) = N G \left( \frac{Q^2}{m N^2} \right).
\end{equation}
In this case again, the $D$ and $G$ functions are \textbf{universal} since they depend on the form of
the $U$ potential, but are independent of the system and of the excitation number
(independent of $N,m,Q$).

From expression (\ref{eq:EF}), one deduces immediately the duality relation
\begin{equation}
\label{eq:dualnN1}
E^{(N)}(m,Q) = \frac{N}{p} E^{(p)} \left( m, \frac{p}{N} Q\right).
\end{equation}
This form is exactly the same as the most general one (\ref{eq:dualpN1}). In this
expression, the same mass appears for systems with $N$ and $p$ particles. The
duality relation leads to link between different excited states of both systems.

Alternatively, one can use the property (\ref{eq:dualgennr}) to obtain many other
possibilities. For example, choosing the value $\beta = N/p$, one has the
alternative duality relation
\begin{equation}
\label{eq:dualnN2}
E^{(N)}(m,Q) = \frac{N}{p} E^{(p)} \left(\frac{N^2}{p^2} m, Q\right).
\end{equation}
In this expression we decide to maintain a one to one correspondence in the spectrum
but for systems with different particle masses.

In (\ref{eq:dualnN1}) and (\ref{eq:dualnN2}), the values of $N$ and $p$ were fixed
numbers; one can consider also $p$ as given but $N = \beta p$ as variable. Thus
$\beta$ is a rational variable such that $\beta p$ is an integer. Then, the last
relation (\ref{eq:dualnN2}) can be recast under an alternative form
\begin{equation}
\label{eq:dualnN3}
E^{(\beta p)}(m,Q) = \beta E^{(p)} \left(\beta^2 m, Q\right).
\end{equation}

\subsection{Two-body interaction only}
\label{sec:no2bodnrv}
In absence of one-body interaction, $U = 0$, we have also interesting conclusions.
Let us choose the special value $a=2/(N-1)$ in the general results of Sect.~\ref{sec:nrUV}.
One has $\sqrt{aN}=N/\sqrt{C_N}$ and $Z(x)=V(x)/N$. Introducing the mean radius $r_0 = y$,
the transcendental equation (\ref{eq:transredn}) reduces to
\begin{equation}
\label{eq:transredn2}
\frac{N Q^2}{m C_N^2} = r_0^3 V'(r_0).
\end{equation}

Defining the $D(x)$ and $G(x)$ function as in (\ref{eq:defD}) and (\ref{eq:defG})
calculated with the potential $V(x)$, one has
\begin{eqnarray}
\label{eq:rofD2}
r_0 & = & D \left( \frac{N Q^2}{m C_N^2} \right), \\
\label{eq:EF2}
E^{(N)}(m,Q) & = & C_N G \left( \frac{N Q^2}{m C_N^2} \right).
\end{eqnarray}
As the previous case studied in Sect.~\ref{sec:no2bodnru}, the $D$ and $G$ functions are \textbf{universal}.

From expression (\ref{eq:EF2}), one deduces immediately the duality relation
\begin{equation}
\label{eq:dualnN3b}
E^{(N)}(m,Q) = \frac{C_N}{C_p} E^{(p)} \left( \frac{p-1}{N-1}m, \frac{p-1}{N-1}
\sqrt{\frac{C_p}{C_N}}Q\right),
\end{equation}
which is identical to the general case (\ref{eq:dualpN4}). Again, one can
use the property (\ref{eq:dualgennr}) to obtain many other possibilities. For example,
choosing the value $\beta = \sqrt{p(N-1)/(N(p-1))}$, one has the alternative simpler
duality relation
\begin{equation}
\label{eq:dualnN4}
E^{(N)}(m,Q) = \frac{C_N}{C_p} E^{(p)} \left(\frac{p}{N} m, \frac{C_p}{C_N} Q \right).
\end{equation}
In this last relation, let us choose $\beta=\sqrt{N/p}$, one arrives at the relation
\begin{equation}
\label{eq:dualnN5}
E^{(N)}(m,Q) = \frac{C_N}{C_p} E^{(p)} \left( m, \frac{p-1}{N-1}\sqrt{\frac{p}{N}} Q \right).
\end{equation}
In this expression, we decide to keep the same mass for systems with $N$ and $p$ particles.
The duality relation leads to a link between different excited states of both systems.
Choosing the value $\beta=C_N/C_p$ in equation~(\ref{eq:dualnN4}), one obtains
the alternative expression
\begin{equation}
\label{eq:dualnN6}
E^{(N)}(m,Q) = \frac{C_N}{C_p} E^{(p)} \left( \frac{(N-1)C_N}{(p-1)C_p} m, Q \right).
\end{equation}
In this expression we decide to maintain a one to one correspondence in the spectrum
but for systems with different particle masses.

\subsection{Link between one and two-body interaction}
\label{sec:no1bodn}

We consider the case of a \textbf{$N$-body system} with particles of mass $m$ and interacting
with a \textbf{two-body interaction $V(x)$ only}. The corresponding binding energy for a state
of excitation number $Q$, $E^{(N)}(m,Q)$, is given by equation (\ref{eq:EF2}).
Now, we consider the case of a $N$-\textbf{body system} with particles of mass $m'$ and interacting
with \textbf{one-body interaction of the form} $U(x) \equiv  c V(x)$ only ($c$ being a constant).
In this case, the corresponding energy for a state of excitation number $Q'$, $\tilde{E}^{(N)}(m',Q')$
is given by the equation
\begin{equation}
\label{eq:dualnrOne}
\tilde{E}^{(N)}(m',Q') = cN G \left( \frac{(Q')^2}{c m' N^2} \right)
\end{equation}
with the same definition of the function $G(x)$.

If one chooses the following link between the parameters,
\begin{equation}
\label{eq:link2mQ}
\frac{NQ^2}{m C_N^2} = \frac{(Q')^2}{c m' N^2} ,
\end{equation}
the arguments of the $G$ function are the same and ${E}^{(N)}$ is expressed as $C_N \tilde{E}^{(N)}/(cN)$.
Therefore we arrive at the following interesting general duality relation (using again (\ref{eq:dualgennr}))
\begin{equation}
\label{eq:dualENtEN}
E^{(N)}(m,Q) = \frac{N-1}{2c} \tilde{E}^{(N)}\left( \frac{\beta^2 (N-1)^2}{4 c N}m, \beta Q \right),
\end{equation}
where $\beta$ is any real parameter.
Choosing $\beta=4c \sqrt{C_N}/(N-1)^2$, one gets
\begin{equation}
\label{eq:dualENtEN1}
E^{(N)}(m,Q) = \frac{N-1}{2c} \tilde{E}^{(N)}\left( \frac{2c}{N-1}m,
               \frac{4c \sqrt{C_N}}{(N-1)^2} Q \right),
\end{equation}
which is a special case of (\ref{eq:dualMNtMN}) applied to a nonrelativistic system.
Choosing $\beta=1$, one has instead
\begin{equation}
\label{eq:dualENtEN2}
E^{(N)}(m,Q) = \frac{N-1}{2c} \tilde{E}^{(N)}\left( \frac{(N-1)^2}{4c \; N}m, Q \right),
\end{equation}
giving a one-to-one correspondence between the spectra of two systems with different
particle masses.
Choosing $\beta=2 \sqrt{c \; N}/(N-1)$, one has rather
\begin{equation}
\label{eq:dualENtEN3}
E^{(N)}(m,Q) = \frac{N-1}{2c} \tilde{E}^{(N)}\left( m, \frac{2 \sqrt{c \; N}}{N-1} Q \right),
\end{equation}
giving a relation between two systems with the same particle mass but for different states
of their spectrum.

\section{Passing from nonrelativistic to ultrarelativistic limits}
\label{sec:passnru}

\subsection{General considerations}
\label{sec:gencons}

For systems submitted to only one type of potential (either one-body potential or two-body
potential) we showed that both the ultrarelativistic limit and the nonrelativistic limit
for the eigenmasses share the property of being expressed in terms of universal functions.
The $F$ function for the ultrarelativistic case and the $G$ function for the nonrelativistic
case are independent of the system and of the excitation quantum numbers but depends only
on the form of the potential under consideration.

If the same form of potential is used in both situations, the expressions of the $F$ function
(see (\ref{eq:defF})) and of the $G$ function (see (\ref{eq:defG})) are not identical so
that the corresponding spectra are quite different.
However, we will show below that if the potentials are different but linked by a certain
relationship, one can arrive at very interesting conclusions.


Let us start from a nonrelativistic problem which is based on a
potential $U(r)$ (no matter one or two-body). One has to calculate
$G(R)$ where $R=R(N,m,Q)$ is a special combination of the parameters
$N$, $m$ and $Q$. This implies to calculate the $x$ value from
\begin{equation}
\label{eq:treq1}
x^3 U'(x) = R
\end{equation}
and then
\begin{equation}
\label{eq:treqG1}
G(R)=\frac{R}{2x^2}+U(x).
\end{equation}
Let us consider now a new nonrelativistic problem with the same value
of $R$ but for a new potential $W(r)$ defined by
\begin{equation}
\label{eq:defW}
W(r)=U(\alpha \sqrt{r}),
\end{equation}
where $\alpha$ is a dimensioned constant (in order that, in both expressions,
$r$ has a dimension of a length while $U$ and $W$ have dimension of energy),
which is undetermined for the moment.

Introducing, instead of $x$, the variable $y$ defined by $x = \alpha \sqrt{y}$.
Starting from (\ref{eq:treq1}), the $y$ quantity results from the transcendental
equation
\begin{equation}
\label{eq:treq2}
y^2 W'(y) = S = \frac{R}{2 \alpha^2},
\end{equation}
while equation (\ref{eq:treqG1}) is transformed into
\begin{equation}
\label{eq:treqG2}
G(R)=\frac{S}{y}+W(y) = F(S).
\end{equation}
In this relation, the $G$ function is the universal function based on potential
$U(r)$ whereas the $F$ function is the universal function based on potential $W(r)$.
This relation allows to propose new duality relations between an ultrarelativistic
and a nonrelativistic treatment. Let us distinguish below the case of one-body
and two-body potentials.

\subsection{One-body interaction only}
\label{sec:dualnruone}

Let us consider a system described by a nonrelativistic treatment based on a
one-body potential $U(r)$ whose binding energy is $E^{(N)}(m,Q)$. We showed in
Sect.~\ref{sec:no2bodnru}  (see equation (\ref{eq:EF}))
\begin{equation}
\label{eq:EF3}
E^{(N)}(m,Q) = N G ( R ) \quad \textrm{with} \quad R = \frac{Q^2}{m N^2}.
\end{equation}
If one uses rather the potential $W(r)$ as in (\ref{eq:defW}), one has, due to property
(\ref{eq:treqG2}),
\begin{equation}
\label{eq:EF4}
E^{(N)}(m,Q) = N F (S) \quad \textrm{with} \quad S = \frac{R}{2 \alpha^2}.
\end{equation}

If we take the special value $S = Q/N$, the value $N F(S)$ represents (see (\ref{eq:massF}))
the ultrarelativistic mass $M_u^{(N)}(Q)$ of the same system but obtained with the
potential $W(r)$. The condition on $S$ and the link between $S$ and $R$ leads
to the condition defining the value of $\alpha$, namely
\begin{equation}
\label{eq:defalpha1}
Q = 2 m N \alpha^2.
\end{equation}

The previous conclusions can be stated as a theorem:
\begin{quote}
If $E^{(N)}(m,Q)$ is the binding energy of a nonrelativistic system governed by
the one-body potential $U(r)$ and if $M_u^{(N)}(Q)$ is the mass of the related
ultrarelativistic system governed by the one-body potential $W(r)$ defined by $W(r) =
U \left( \sqrt{Qr/(2 m N)} \right)$, then one has the general property
$M_u^{(N)}(Q) = E^{(N)}(m,Q)$.
\end{quote}
From its definition, the potential $W(r)$ depends on $m$ so that the notation
$M_u^{(N)}(Q)$ which we have used up to now, depends indirectly on $m$, as imposed
by the theorem.

\subsection{Two-body interaction only}
\label{sec:dualnrutwo}

Let us now consider a system described by a nonrelativistic treatment based on a
two-body potential $V(r)$ whose binding energy is $E^{(N)}(m,Q)$. We showed in
Sect.~\ref{sec:no2bodnrv}  (see equation (\ref{eq:EF2}))
\begin{equation}
\label{eq:EF5}
E^{(N)}(m,Q) = C_N G \left( R \right)  \quad \textrm{with} \quad R = \frac{N Q^2}{m C_N^2}.
\end{equation}
If one uses rather the potential $W(r)$ as in (\ref{eq:defW}), one has, due to property
(\ref{eq:treqG2}),
\begin{equation}
\label{eq:EF6}
E^{(N)}(m,Q) = C_N F (S)  \quad \textrm{with} \quad S = \frac{R}{2 \alpha^2}.
\end{equation}

If we take the special value $S = 2Q/((N-1)\sqrt{C_N})$, the value $C_N F(S)$
represents (see (\ref{eq:massF2})) the ultrarelativistic mass $M_u^{(N)}(Q)$ of the
same system but obtained with the potential $W(r)$. The condition on $S$ and the
link between $S$ and $R$ leads to the condition defining the value of $\alpha$, namely
\begin{equation}
\label{eq:defalpha1b}
Q = 2m \sqrt{C_N} \alpha^2.
\end{equation}

In this case again, one can state the following theorem:
\begin{quote}
If $E^{(N)}(m,Q)$ is the binding energy of a nonrelativistic system governed
by the two-body potential $V(r)$ and if $M_u^{(N)}(Q)$ is the mass of the related
ultrarelativistic system governed by the two-body potential $W(r)$ defined by $W(r) =
V \left( \sqrt{Qr/(2m \sqrt{C_N})} \right)$, then one has the general property
$M_u^{(N)}(Q) = E^{(N)}(m,Q)$.
\end{quote}
We can mention the same remark as before concerning the indirect $m$ dependence of $M_u^{(N)}(Q)$.

\section{Applications}
\label{sec:applic}

All the duality relations presented in the previous sections are exact for the AFM solutions
of quantum systems.  We can wonder to what extent these constraints are satisfied for the
corresponding  exact solutions. In this section, we will examine the relevance of some duality
relations for  genuine solutions of particular systems.

We will not test all the duality relations which are presented in this paper (some of them are
consequences of others and do not need a second check), but we will look at the cases of two-body
interactions only ($U=0$) and one-body interactions only ($V=0$), as well as nonrelativistic and
ultrarelativistic kinematics. In a first study, we will consider nonrelativistic systems, for
which relation (\ref{eq:dualgennr}) is specially important and will be used intensively.

\subsection{Nonrelativistic systems}
\label{testnr}

\subsubsection{General considerations}
\label{gctestnr}

Let us call $\epsilon^{(N)}(m;\{n_i,l_i\})$ the \textbf{exact eigenenergy} of the $\{n_i,l_i\} =
\{n_1,l_1,n_2,l_2,\ldots,n_{N-1},l_{N-1}\}$ state of the Hamiltonian
\begin{equation}
\label{eq:HnrN}
H = \sum_{i=1}^N \frac{\bm p_i^2}{2m} + \sum_{i<j=1}^N V(|\bm r_i - \bm r_j |),
\end{equation}
while $\epsilon(m;n,l) = \epsilon^{(2)}(m;\{n,l\})$ gives the exact spectrum of the
corresponding two-body problem. $E^{(N)}(m;Q^{(N)})$ and $E(m;Q)$ are the \textbf{approximate
AFM eigenenergies} of the same states. The duality relations are exactly true for the $E$
values but not for the $\epsilon$ values in general. In this Sect.~\ref{testnr}, we will consider
only systems of boson-like particles, in order that the ground state is given by
$\{n_i,l_i \} = \{ 0,0 \}$.

Let us assume that we are able to obtain, for each set of parameters $(m,n,l)$, the value
$\epsilon(m;n,l)$ for a two-body system. Nowadays, it is not difficult to get numerical 
solutions for such a problem. We can compute the corresponding
value $E(m;Q)$ analytically or, if not, numerically. Equating both values defines the $Q$
value that leads to the exact value in the AFM expression. Of course, this value depends
on $(n,l)$ but also, most of the time, on $m$. We note this value $Q(m;n,l)$ which is thus
defined by the formal equality
\begin{equation}
\label{eq:defQmnl}
E(m;Q(m;n,l)) = \epsilon(m;n,l).
\end{equation}
For the harmonic oscillator (HO), the $Q$ value is $m$-independent and is equal to $2n+l+3/2$, whereas for
the Coulomb potential the $Q$ value is still $m$-independent and is equal to $n+l+1$.
We showed that for a lot of potentials, the $m$-dependence of $Q$ is not
crucial but that a good dependence in $(n,l)$ is of major importance \cite{bsb10b}. For the rest of
this section, we suppose that we adopt a form of $Q$ depending on $n$ and $l$, but
not on $m$, which gives quite good results. Consequently the relation (\ref{eq:defQmnl})
does not remain an equality but holds only approximately
\begin{equation}
\label{eq:defQnl}
E(m;Q(n,l)) \approx \epsilon(m;n,l).
\end{equation}

The duality relation (\ref{eq:dualgennr}) allows to express the AFM energy in terms of
the energy of the ground state but for a different mass. Denoting $Q(0,0)=Q_2$, one
has explicitly
\begin{equation}
\label{eq:dualQnl}
E(m;Q(n,l)) = E(\bar m(m;n,l);Q_2),
\end{equation}
with the definition of the $\bar m(m;n,l)$ mass
\begin{equation}
\label{eq:defmnl}
\bar m(m;n,l) = \frac{Q_2^2}{Q(n,l)^2} m.
\end{equation}
These equations are equalities. Using them, with the approximate relation (\ref{eq:defQnl}),
one arrives at the approximate duality condition concerning the exact states
\begin{equation}
\label{eq:dualepsnl}
\epsilon(m;n,l) \approx \epsilon(\bar m(m;n,l);0,0).
\end{equation}

Let us define the $f$ function by
\begin{equation}
\label{eq:deffuncf}
f(m) = \epsilon(m;0,0).
\end{equation}
This function is \textbf{universal} in the sense that it depends only on the
form of the potential $V(r)$ and it can be computed once for all. The
relation (\ref{eq:dualepsnl}) can then be recast under the form
\begin{equation}
\label{eq:epsnlfm}
\epsilon(m;n,l) \approx f(\bar m(m;n,l)).
\end{equation}
The conclusion is very strong. It means that the entire spectrum of a given
two-body system can be obtained, at least approximatively, by using in an
universal function, $f(m)$, the various arguments $\bar m(m;n,l)$ given by
(\ref{eq:defmnl}). The quality of the results depends of course on the
ability for the function $Q(n,l)$ to reproduce the exact results in a
satisfactory way.
The test of duality relation (\ref{eq:epsnlfm}) in a realistic case is the
subject of Sect.~\ref{test2bnr}.

The relation (\ref{eq:dualgennr}) being valid for an arbitrary number of
particles, all the demonstrations that we have developed above for
the two-body problem apply as well for the $N$-body problem. One can
define formally a principal quantum number $Q^{(N)}(m;\{n_i,l_i\})$ through
an equality similar to (\ref{eq:defQmnl}). The crucial hypothesis is that 
this value depends only slightly on the $m$ parameter and can be very
well approximated by a simpler function $Q^{(N)}(\{n_i,l_i\})$. Denoting
this time the principal quantum number for the ground state as $Q_N =
Q^{(N)}(\{0,0\})$ the equivalent of equation (\ref{eq:dualQnl}) now writes
\begin{equation}
\label{eq:dualQnlN} 
E^{(N)}(m;Q^{(N)}(\{n_i,l_i\})) = E(\bar m(m;\{n_i,l_i\});Q_N),
\end{equation}
with the definition of the $\bar m(m;\{n_i,l_i\})$ mass
\begin{equation}
\label{eq:defmnlN}
\bar m(m;\{n_i,l_i\}) = \frac{Q_N^2}{(Q^{(N)}(\{n_i,l_i\}))^2} m.
\end{equation}
Thus, one has an approximate duality relation concerning the $N$-body spectrum
in term of the ground state of the same system for another mass
\begin{equation}
\label{eq:dualepsnlN}
\epsilon^{(N)}(m;\{n_i,l_i\}) \approx \epsilon^{(N)}(\bar m(m;\{n_i,l_i\});\{0,0\}).
\end{equation}
The test of duality relation (\ref{eq:dualepsnlN}) in a realistic case for
$N=3$ will be discussed in Sect.~\ref{test3bnr}.

Finally, a link can be found between the $N$-body and
the 2-body systems. Since a link between excited states and the ground state has
been proposed above, it is sufficient to search for a relationship between ground
states. In order to do that, let us apply relation (\ref{eq:dualnN4}) for $p=2$
leading to (with again the use of (\ref{eq:dualgennr}))
\begin{equation}
\label{eq:linkN2gs}
E^{(N)}(m;Q) = C_N E(2m/N;Q/C_N) = C_N E(2 \alpha^2 m/N;\alpha Q/C_N),
\end{equation}
where $\alpha$ is an arbitrary real parameter.
Let us choose the value $m=\bar m(m;\{n_i,l_i\})$, $Q=Q_N$ and $\alpha = C_N Q_2/Q_N$
in the previous equation. One gets
\begin{equation}
\label{eq:linkN2gs2}
E^{(N)}(\bar m(m;\{n_i,l_i\});Q_N) = C_N E(M(m;\{n_i,l_i\});Q_2),
\end{equation}
with the definition of the $M$ mass
\begin{equation}
\label{eq:defMassnl}  
M(m;\{n_i,l_i\}) = \frac{2m}{N} \left( \frac{C_N Q_2}{Q^{(N)}(\{n_i,l_i\})} \right)^2.
\end{equation}

If the AFM energies give a good approximation of the exact results, one can expect
that $E^{(N)}(\bar m(m;\{n_i,l_i\});Q_N) \approx \epsilon^{(N)}(\bar m(m;\{n_i,l_i\});\{0,0\})$
and $E(M(m;\{n_i,l_i\});Q_2) \approx \epsilon(M(m;\{n_i,l_i\});0,0) =
f(M(m;\{n_i,l_i\})$. Owing to the relations (\ref{eq:dualepsnlN}) and (\ref{eq:linkN2gs2}),
one arrives at the very important result
\begin{equation}
\label{eq:dualepsnlN2}
\epsilon^{(N)}(m;\{n_i,l_i\}) \approx C_N f(M(m;\{n_i,l_i\}).
\end{equation}
The conclusion of this relation is even stronger than (\ref{eq:epsnlfm}). Equation
(\ref{eq:dualepsnlN2}) proves that \textbf{the whole spectrum of all systems} can be
obtained \textbf{approximatively} by the calculation of a universal function, $f(m)$, corresponding
to the ground state of the 2-body system with the same potential, and for arguments
$M(m;\{n_i,l_i\})$ given by (\ref{eq:defMassnl}). Getting $f(m)$ is a very easy task. 
Solving the 2-body system can be performed with a great accuracy for any potential
(let us recall that this potential must not depend on $m$ and $N$); moreover, obtaining
the ground state energy is free from possible numerical complications arising
for excited states. Let us note also that the duality relation (\ref{eq:dualepsnlN2}),
which is obviously an approximation, concern the exact eigenvalues only; the AFM
values, which were very convenient intermediate quantities in our demonstration, have
completely disappeared.
Testing the link between the ground states of a 2-body and a 3-body system is the
subject of Sect.~\ref{test23bnr}.

Let us point out an additional comment. All what we did  for the ground state of a two-body
system can be reproduced identically for any other excited state $(n_0,l_0)$. It is possible
to introduce a universal function $f_{n_0,l_0}(m)$ built on this state. Then, (\ref{eq:dualepsnlN2})
can be expressed in term of the $f_{n_0,l_0}$ function, but with a new argument given by
(\ref{eq:defMassnl}) in which $Q_2$ has been replaced by $Q(n_0,l_0)$.

The scaling laws or the dimensional analysis allow generally to rewrite a Hamiltonian on
a reduced form, easier to study. The eigenenergies can then be expressed in terms of a
dimensioned parameter (an energy scale) and other dimensionless parameters which depends
on the physical quantities of the problem (particles masses, length scales, etc.) but
not on the quantum numbers of the corresponding eigenstates. That is why equations
(\ref{eq:epsnlfm}) and (\ref{eq:dualepsnlN2}) are both of different nature than the
usual scaling laws. 

\subsubsection{Testing a two-body system}
\label{test2bnr}

\begin{figure}[htb]
\includegraphics*[height=4cm]{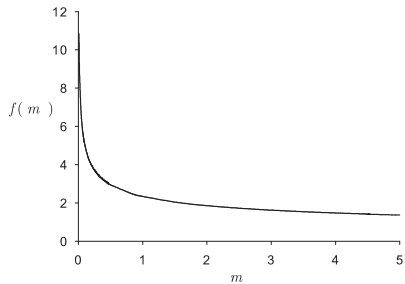}
\caption{Universal function $f(m)$ as a function of the mass $m$ for
Hamiltonian~(\ref{eq:H2blin}).}
\label{fig1}
\end{figure}

In this section, we test the duality relation (\ref{eq:epsnlfm}) for two particles
interacting with a pure linear potential $V(r)=r$, so that the corresponding
Hamiltonian in reduced variables is simply
\begin{equation}
\label{eq:H2blin}
H = \frac{\bm p^2}{m}+r.
\end{equation}
The first thing to do is to obtain the ground state energy as a function of the mass
in order to plot the universal function $f(m)$ for the linear potential. The
numerical value has been calculated with a great accuracy using the so-called
Lagrange mesh method \cite{sema01}. The form of the $f(m)$ function is shown
in Figure~\ref{fig1}. This function decreases very rapidly from infinity at the
limit $m \to 0$ to around 2 for $m=1$; then, it decreases very slowly to zero
for large values of $m$.

Now we calculate, still using the Lagrange mesh method, the spectrum of a system
with mass $m=4$, in order to match the results of \cite{bsb08a}. 
We think that limiting ourselves to states with $0 \leq n \leq 3$
and $0 \leq l \leq 3$ is enough for our purpose; the corresponding eigenenergies
$\epsilon(m=4;n,l)$ are given in the first line of Table \ref{Tab1}.

In order to calculate the effective masses $\bar m(4;n,l)$ as defined in (\ref{eq:defmnl}),
we need a prescription for the choice of the principal quantum number. We study here
two possibilities: the HO prescription $Q(n,l) = 2n+l+3/2$ and the
improved value proposed in \cite{bsb08a} $Q(n,l) = 1.789n + l + 1.375$. For both
prescriptions, we calculate the effective masses (\ref{eq:defmnl}) and insert them
in the $f(m)$ function to obtain values $f(\bar m(4;n,l))$ which are ultimately
compared to the exact results $\epsilon(4;n,l)$. The approximate values $f(\bar m(4;n,l))$
are reported in the second line of Table \ref{Tab1} for the HO prescription whereas
the third line of the same Table shows the results with the improved formula.
Note the the ground state is always exactly computed by definition.

\begin{table}[ht]
\begin{center}
\caption{Values of $\epsilon(m=4;n,l)$ for various approximations. For each set $(n,l)$, 
the first line is the result obtained by numerical integration $\epsilon(4;n,l)$,
the second and third lines are the results coming from duality relation $f(\bar m(4;n,l))$
with the HO prescription $Q(n,l)=2n+l+3/2$ and the improved prescription
$Q(n,l)=1.789n + l + 1.375$, respectively.
Between parenthesis is also indicated the deviation (\%) of the approximate
result as compared to the exact ones.
}
\label{Tab1}
\begin{tabular}{lllll}
\hline\hline
 $l$ & $\epsilon(4;0,l)$ & $\epsilon(4;1,l)$ & $\epsilon(4;2,l)$ & $\epsilon(4;3,l)$ \\
\hline
 0 & \bf{1.473} &  \bf{2.575} &  \bf{3.478} &  \bf{4.275} \\
   & 1.473 (0) &  2.591 (0.6) &  3.502 (0.7) &  4.307 (0.7) \\
   & 1.473 (0) &  2.567 (0.3) &  3.461 (0.5) &  4.251 (0.5) \\
 1 & \bf{2.117} &  \bf{3.077} &  \bf{3.911} &  \bf{4.665} \\
   & 2.071 (2.2) &  3.064 (0.4) &  3.915 (0.1) &  4.682 (0.3) \\
   & 2.120 (0.1) &  3.083 (0.2) &  3.913 (0.05) &  4.662 (0.06) \\
 2 & \bf{2.676} &  \bf{3.546} &  \bf{4.327} &  \bf{5.046} \\
   & 2.591 (3.2) &  3.502 (1.2) &  4.307 (0.5) &  5.042 (0.08) \\
   & 2.680 (1.5) &  3.559 (0.4) &  4.339 (0.3) &  5.055 (0.2) \\
 3 & \bf{3.182} &  \bf{3.989} &  \bf{4.728} &  \bf{5.416} \\
   & 3.064 (3.7) &  3.915 (1.8) &  4.682 (1.0) &  5.390 (0.5) \\
   & 3.186 (0.1) &  4.005 (0.4) &  4.746 (0.4) &  5.434 (0.3) \\
\hline\hline
\end{tabular}
\end{center}
\end{table}

From the table one sees that the duality relation~(\ref{eq:epsnlfm}) is
very well satisfied in any case. The very simple HO prescription is less good
(specially for large $l$ values) but remains acceptable with discrepancies 
around 1\%. The improved prescription, which raises the degeneracy, is very impressive (even for
large $l$ values) with an average precision of the order of 0.5\%.

We checked also the validity of this duality relation for other types of potential 
(square root [$\sqrt{r^2+a}$], funnel [$-a/r+b r$])
 and find that the accuracy is always of the order of 1\% or better.
Thus, we think that the studied duality relation is well adapted for the spectrum
of a two-body problem.

\subsubsection{Testing a three-body system}
\label{test3bnr}

In this section, we test the duality relation (\ref{eq:dualepsnlN}) for three particles,
with a mass $m$, interacting with a pure linear potential $V(r)=r$, so that the corresponding
Hamiltonian in reduced variables is simply
\begin{equation}
\label{eq:h3lin}
H = \sum_{i=1}^3 \frac{\bm p_i^2}{2 m} + \sum_{i<j = 1}^3 |\bm r_i - \bm r_j |.
\end{equation}
To compute the solutions of this Hamiltonian, we use 
a variational method relying on the expansion of trial states with a HO
basis \cite{hobasis}. We can write
\begin{equation}
\label{hotrial}
|\psi\rangle=\sum_{B=0}^{B_\textrm{max}}\sum_{q(B)} |\phi(B,q(B))\rangle,
\end{equation}
where $B=2(n_1+n_2)+(l_1+l_2)$ characterizes the band number of the basis state and where $q$
summarizes all the quantum numbers of the state (which can depend on $B$). This
procedure is specially interesting since the eigenstates of $H$ are
expanded in terms of AFM eigenstates (up to a length scale
factor) \cite{silv10,bsb10b}. In practice, a relative accuracy better than $10^{-4}$ is reached with
$B_\textrm{max}=20$. Such results are denoted ``exact" in the following. Some exact eigenvalues 
$\epsilon$ of (\ref{eq:h3lin}) for $m=2$ are presented in Table~\ref{Tab2} with the quantum numbers 
$\{n_1,l_1,n_2,l_2\}$ and the $B$ value of the main component in the expansion (\ref{hotrial}).

In order to compute the effective mass (\ref{eq:defmnlN}), we need a prescription for the principal
quantum number. We consider two formulas. The first one is the HO result: $Q_{\textrm{HO}}=
2(n_1+n_2)+(l_1+l_2)+3$. The second one is given by $Q_{\textrm{WKB}}=\frac{\pi}{\sqrt{3}}
(n_1+n_2)+(l_1+l_2)+3$. This prescription comes from the fact that the ratio between the coefficients
of $n$ and $l$ for a nonrelativistic particle in a linear potential is predicted to be $\pi/\sqrt{3}$
by a WKB method \cite{brau00} and the AFM \cite{bsb08a}. The set of quantum numbers used for a state
is fixed by the quantum numbers of the main component in the expansion (\ref{hotrial}). It is then
possible to compute the energy of an excited state from (\ref{eq:defmnlN}) and (\ref{eq:dualepsnlN}).
The corresponding values denoted $\epsilon(Q_{\textrm{HO}})$ and $\epsilon(Q_{\textrm{WKB}})$ are
presented in table~\ref{Tab2}. Note the the ground state is always exactly computed by definition.

\begin{table}[ht]
\begin{center}
\caption{Some exact eigenvalues $\epsilon$ of (\ref{eq:h3lin}) for $m=2$ with the quantum numbers 
$\{n_1,l_1,n_2,l_2\}$ and the $B$ value of the main component in expansion (\ref{hotrial}).
The brackets indicate that two components are equally present with $n_1 \leftrightarrow n_2$ 
and $l_1 \leftrightarrow l_2$. Energies predicted by (\ref{eq:defmnlN}) and (\ref{eq:dualepsnlN})
are given in columns $\epsilon(Q_{\textrm{HO}})$ and $\epsilon(Q_{\textrm{WKB}})$ following
the prescription chosen for the principal quantum numbers (see text). 
Between parenthesis is also indicated the deviation (\%) of the approximate
result as compared to the exact ones.
}
\label{Tab2}
\begin{tabular}{cccll}
\hline\hline
$B$ & $n_1,l_1,n_2,l_2$ & $\epsilon$ & $\epsilon(Q_{\textrm{HO}})$ & $\epsilon(Q_{\textrm{WKB}})$ \\
\hline
0 & 0,0,0,0   & \bf{4.867} & 4.867       & 4.867       \\
1 & [0,1,0,0] & \bf{5.934} & 5.896 (0.7) & 5.896 (0.7) \\
2 & [1,0,0,0] & \bf{6.704} & 6.842 (2.1) & 6.671 (0.5) \\
  & 0,1,0,1   & \bf{6.846} & 6.842 (0.1) & 6.842 (0.1) \\
  & [0,2,0,0] & \bf{6.874} & 6.842 (0.5) & 6.842 (0.5) \\ 
3 & [1,1,0,0] & \bf{7.608} & 7.726 (1.6) & 7.566 (0.6) \\ 
  & 1,0,0,1   & \bf{7.702} & 7.726 (0.3) & 7.566 (1.8) \\ 
  & [0,2,0,1] & \bf{7.854} & 7.726 (1.6) & 7.726 (1.6) \\
4 & [2,0,0,0] & \bf{8.309} & 8.562 (3.0) & 8.256 (0.6) \\
  & 1,1,0,1   & \bf{8.391} & 8.562 (2.0) & 8.410 (0.2) \\
  & [1,2,0,0] & \bf{8.426} & 8.562 (1.6) & 8.410 (0.2) \\
  & 0,1,1,1   & \bf{8.572} & 8.562 (0.1) & 8.410 (1.9) \\
  & 0,2,0,2   & \bf{8.707} & 8.562 (1.7) & 8.562 (1.7) \\
\hline\hline
\end{tabular}
\end{center}
\end{table}

From this table one sees that the duality relation~(\ref{eq:defmnlN})-(\ref{eq:dualepsnlN}) is
very well satisfied in any case. The very simple $Q_{\textrm{HO}}$ prescription already gives
good results, but the $Q_{\textrm{WKB}}$ prescription raises partly the degeneracy and 
improves appreciably the global agreement. 

\subsubsection{From two- to three-body ground states}
\label{test23bnr}

Let us consider a system of particles interacting via two-body forces
and let us assume that the principal quantum 
number of the ground state can be written $Q_N=\rho (N-1)$ with $\rho$ independent of $N$. 
It is true for the $N$-body HO with $\rho=3/2$ (see~(\ref{eq:princnumoh})). In this case, using
(\ref{eq:linkN2gs2}) and (\ref{eq:defMassnl}), and assuming that the AFM solution is a good 
approximation of the true solution, we can write
\begin{equation}
\label{eq:gs2gs3}
\epsilon^{(N)}(m;\{0,0\}) \approx C_N \epsilon(Nm/2;0,0).
\end{equation}
We have then an approximate link between the ground state of $N$ particles 
interacting via a two-body potential and the ground state of the corresponding two-body
system. We have checked that, for $N=3$, relation (\ref{eq:gs2gs3}) is satisfied with an error 
less that 1\% for $V(x)=a x$, around 1\% for $V(x)=-a/x+b x$ and around 6\% for $V(x)=-a/x$
(these potentials deviate more and more from a quadratic one).

The three tests performed in sections \ref{test2bnr}, \ref{test3bnr} and \ref{test23bnr}
prove that the most general formulation of the duality relations, as expressed by
(\ref{eq:dualepsnlN2}), is indeed satisfied with a rather good accuracy of the order of a
few percent, at least for $N=2,3$.

\subsection{An ultrarelativistic system}
\label{sec:applic1}

Let us consider the following ultrarelativistic Hamiltonian
\begin{equation}
\label{eq:urhotest}
H = \sum_{i=1}^N \sqrt{\bm p_i^2} + a \sum_{i=1}^N |\bm r_i - \bm R|,
\end{equation}
which presents some interest for hadronic physics \cite{barlnc} and 
whose AFM eigenmasses are given by $M^{(N)}_u(Q) = \sqrt{4 N a Q}$. In \cite{silv10},
it has been shown that the accuracy of this formula can be improved as compared to the pure
HO prescription in the case $N=3$ by the determination in (\ref{eq:princnumod}) of:
\begin{itemize}
  \item The $\gamma$ term by using an accurate variational solution for the ground state.
  \item The ratio $\alpha_i/\beta_i$ by using the characteristics of a WKB solution for an ultrarelativistic particle in a linear potential \cite{brau00}.
\end{itemize}
The same method can be used in the case $N=2$. Finally, we obtain
\begin{eqnarray}
\label{M2test}
M^{(2)}_u&\approx& \sqrt{8 a Q^{(2)}}\quad \textrm{with}\quad Q^{(2)}=\frac{\pi}{2} n + l +
\frac{4}{\pi},\\
\label{M3test}
M^{(3)}_u&\approx& \sqrt{\frac{32}{\pi} a Q^{(3)}}\quad \textrm{with}\quad\ Q^{(3)}=
\frac{\pi}{2} (n_1 + n_2) + l_1 + l_2 + 3.
\end{eqnarray}
The relative errors are below 2\% for (\ref{M2test}) and below 1\% for (\ref{M3test}) (at
least for quantum numbers such that $n \le 3$ and $l \le 3$). So,we can consider that these
formulas are very close to the exact solutions. 

We can now check if relation (\ref{eq:dualuN1}) is, at least approximately, verified with
$N=3$ and $p=2$ for the (nearly) exact solutions of this system. A priori, we could expect
that $M^{(3)}_u(Q^{(3)}) \approx 3/2\ M^{(2)}_u(2Q^{(3)}/3)$. The calculation gives 
\begin{equation}
\label{eq:chM3M2ur}
\frac{3}{2} M^{(2)}_u\left(\frac{2Q^{(3)}}{3}\right) \approx \frac{3}{2}
\sqrt{8 a \frac{2Q^{(3)}}{3}} =  \sqrt{12 a Q^{(3)}}.
\end{equation}
This must be compared with (\ref{M3test}). Since $32/\pi\approx 10.2$, the relative error
is around 15\%.

But, we can also hope that $M^{(2)}_u(Q^{(2)}) \approx 2/3\ M^{(3)}_u(3Q^{(2)}/2)$. The
calculation gives 
\begin{equation}
\label{eq:chM2M3ur}
\frac{2}{3} M^{(3)}_u\left(\frac{3Q^{(2)}}{2}\right) \approx \frac{2}{3}
\sqrt{\frac{32}{\pi} a \frac{3Q^{(2)}}{2}} = 
\sqrt{\frac{64}{3\pi} a Q^{(2)}}.
\end{equation} 
This must be compared with (\ref{M2test}). Since $64/(3\pi)\approx 6.8$, the relative
error is also around 15\%. In these cases, we can consider that the relevance of the
duality relation is demonstrated.

\subsection{An ultrarelativistic-nonrelativistic duality}
\label{sec:applic2}

The eigensolutions of these two two-body Hamiltonians
\begin{eqnarray}
\label{thnur1}
H_{n}&=&\frac{\bm p^2}{m}+a r^2, \\
\label{thnur2}
H_{u}&=&2 \sqrt{\bm p^2}+b r, 
\end{eqnarray}
are given respectively by
\begin{eqnarray}
\label{tenur1}
E_n(Q_n)&=&\sqrt{\frac{4 a}{m}}Q_n , \quad \textrm{with}\quad Q_n=2 n + l + \frac{3}{2}, \\
\label{tenur2}
M_u(Q_u)&\approx&\sqrt{8 b Q_u}\quad \textrm{with}\quad Q_u=\frac{\pi}{2} n + l + \frac{4}{\pi}.
\end{eqnarray}
Relation (\ref{tenur1}) is exact while the relative accuracy of (\ref{tenur2}) is around 1-2\%. 
If we impose $Q_n=Q_u=Q$, the duality relation presented in Sect.~\ref{sec:passnru} is verified
by setting $2 b m=a Q$ as shown above. 

Let us examine what becomes this property for the genuine solutions. If we assume that $E_n$
is known, we can compute the exact solutions of $H_{u}$ with $b=a Q_n/ (2 m)$. For $n\le 3$ and
$l\le 3$, the relative error compared to (\ref{tenur2}) is below 10\%. On the contrary, if we assume
that $M_u$ is known, we can compute the exact solutions of $H_{n}$ with $a=2 b m /Q_u$. For
$n\le 3$ and $l\le 3$, the relative error compared to (\ref{tenur1}) can then reach 38\%. Following
the physical situation considered, the accuracy of this kind of duality relation can
strongly vary. But, in some cases, the error could be quite small.

\section{Conclusions}
\label{sec:conclu}

In this paper, we investigate the possible relationships between the energies of particular states for a
given system and the energies of other states for another system. We call these relationships ``duality
relations". Systems with arbitrary number of particles are considered but in a well defined framework. 
The basic hypotheses are the following ones:
\begin{itemize}
  \item All the particles of the studied systems are identical.
  \item We do not take into account internal degrees of freedom for particles and, for
convenience, we focus only on bosons in applications (the case of fermions can be dealt with as well, but
it is more involved and is not studied in this work).
  \item We consider the kinetic energy operator either of a nonrelativistic form or of
a semirelativistic form.
  \item We limit ourselves to one-body and two-body types of potentials.
  \item The potentials do not lead to coupled channel equations, so that we are faced
with a Schr\"{o}dinger equation for a nonrelativistic kinetic energy, and with a spinless
Salpeter equation for a semirelativistic kinetic energy.
  \item The potentials entering the formalism do not depend on the mass $m$ nor on the
number of particles $N$.
\end{itemize}
Despite the limitations, the number of systems which can be concerned by this work 
is quite large and may represent a lot of different physical situations.

As an intermediate tool, we introduce the auxiliary field method (AFM) to get approximate
values of the eigenenergies. The important parameters are the mass $m$ and the principal
quantum number $Q(\{n_i,l_i\})$, function of the various radial quantum numbers $n_i$ and
orbital quantum numbers $l_i$. The form of this function is not given by the theory and
is left to the cleverness of the physicist. Usually, the harmonic oscillator prescription
(\ref{eq:princnumoh}) is a good starting point, but a more elaborated formula, such
as the one given  by (\ref{eq:princnumod}), can improve substantially the quality of the
results. The AFM provides an expression $E^{(N)}(m;Q)$ of the eigenenergies 
(or eigenmasses for a semirelativistic kinematics) in terms of these
parameters. In a series of papers \cite{bsb08a,bsb08b,AFMeigen,bsb09a,Sem09a,bsb09c,silv10,bsb10b},
we proved that, the AFM approximation
can be calculated analytically for many systems and leads to encouraging results.

A duality relation is, at the beginning, a link between AFM approximations
$E^{(N)}(m;Q)$ and $E^{(p)}(m';Q')$. The case $p=2$
is particularly important because the corresponding duality relation allows to obtain
the spectrum of a $N$-body system in terms of the spectrum of a 2-body system, a much more
favorable situation.
We obtained very general duality relations when only one-body or two-body potentials are
present. More interesting conclusions can be drawn in the ultrarelativistic limit ($m=0$)
or in the nonrelativistic limit. In these particular cases, we showed that the AFM
results can be expressed in term of a unique universal function for a given potential 
and for a given kinematics. 
This crucial feature has a lot of important consequences for the duality relations.
We proved for example that the spectrum of a nonrelativistic system governed by a
two-body potential (the same property holds for a one-body potential) $V(r)$ is the
same as the spectrum of an ultrarelativistic system governed by a two-body potential
$W(r)=V(\alpha \sqrt{r})$ provided that the parameter $\alpha$ is chosen adequately.
The nonrelativistic kinematics allows additional sympathetic features. In particular,
it was shown that the spectrum of such a $N$-body system can be obtained from the ground
state of the corresponding two-body system.

The duality relations presented here are exact for the AFM approximation of the
eigenenergies. Assuming that this approximation leads to results close to the exact
ones, the duality relations can also be applied to the exact states. This means that
we have an \textbf{approximate} link between the exact energies of a 
$N$-body system $\epsilon^{(N)}(m;\{n_i,l_i\})$
and the exact energies of a $p$-body system $\epsilon^{(p)}(m';\{n'_i,l'_i\})$. The
case $p=2$ is specially interesting because nowadays we are able to get easily the 
spectrum of a two-body system numerically with a great accuracy. 

In particular, for a nonrelativistic system, 
we obtained a very important conclusion. The whole spectrum
of any system (arbitrary $N$ value) can be calculated, approximatively, through a
universal function $f(m)$ which is nothing else than the dependence of the ground state
of the two-body system (with the same potential) on the mass. This function is very
easy to obtain numerically in any circumstances. The duality relation is expressed
by equation (\ref{eq:dualepsnlN2}). It is sufficient for obtaining the searched value
$\epsilon^{(N)}(m;\{n_i,l_i\})$ to calculate the universal function for an argument
$M(m;\{n_i,l_i\})$ given by (\ref{eq:defMassnl}).

The important point that we want to stress is the following one. The duality relations
are exact for the AFM approximations of the eigenenergies while they are only
approximately true for the exact eigenenergies (the quality probably deteriorates
with increasing values of $N$). This drawback is compensated by the fact that the
corresponding duality relations concern the exact eigenenergies without any reference
to the AFM expressions which were only an intermediate tool.
Thus, the consequences of duality relations would remain true independently of 
the AFM approximations. We propose to apply all the duality relations presented above
for the exact eigenvalues, but only in an approximate way.

The duality relations have been tested in a number of interesting cases.
In particular, using a
linear potential for $N=2$ and 3, we showed that they are fulfilled with an accuracy
of the order of 1-4\%. For other types of potentials, for larger values
of $N$ and quantum numbers $\{n_i,l_i\}$, the quality is likely poorer, but we think
that these duality relations are nevertheless able to give valuable informations
concerning complicated systems which are not easy to be dealt with, analytically or numerically.

\appendix

\section{Tests of duality relations for AFM energies}
\label{sec:testsd}

In order to test the duality relations presented above, we can use the semirelativistic $N$-body 
harmonic oscillator studied in \cite{silv10}
\begin{equation}
\label{eq:urho}
H = \sum_{i=1}^N \sqrt{\bm p_i^2+m^2} + k \sum_{i=1}^N |\bm r_i - \bm R|^2 +
    \rho \sum_{i<j = 1}^N |\bm r_i - \bm r_j |^2. 
\end{equation}
The AFM solutions, for the general, the ultrarelativistic and the nonrelativistic cases respectively,
are given by 
\begin{eqnarray}
\label{tho1}
M^{(N)}&=&\frac{2 N m}{\sqrt{3 Y_N}}\left[ \frac{1}{G_{-}(Y_N)} + G_{-}(Y_N)^2 \right]
\quad \textrm{with} \quad Y_N=\frac{4 m^2}{3}\left( \frac{2 N^2}{(k+\rho N)Q^2} \right)^{2/3}, \\
\label{tho2}
M^{(N)}_u&=&\frac{3}{2}\left[ 2 N (k+\rho N)Q^2 \right]^{1/3}, \\
\label{tho3}
E^{(N)}&=&\sqrt{\frac{2}{m}(k+\rho N)}Q.
\end{eqnarray}
Let us mention that the last formula is an exact solution. The function $G_{-}(Y)$ is the root of a 
quartic polynomial whose explicit form is given in \cite{bsb09c} for instance. Using these formulas,
it is easy to check the relations (\ref{eq:dualpN1}), (\ref{eq:dualpN4}), (\ref{eq:dualMNtMN}),
(\ref{eq:dualuN1}), (\ref{eq:dualuN2}), (\ref{eq:dualMNtMNu}), (\ref{eq:dualgennr}),
(\ref{eq:dualnN1})-(\ref{eq:dualnN3}), (\ref{eq:dualnN3b})-(\ref{eq:dualnN6}),
(\ref{eq:dualENtEN})-(\ref{eq:dualENtEN3}).

The semirelativistic harmonic oscillator with one variable, studied in \cite{bsb09c},
has the following form
\begin{equation}
\label{eq:urhos}
H = \sigma \sqrt{\bm p^2+m^2} + a r^2. 
\end{equation}
The AFM solutions, for the general and the ultrarelativistic cases respectively,
are given by 
\begin{eqnarray}
\label{tho4}
M^{(s)}&=&\frac{2 \sigma m}{\sqrt{3 Y_s}}\left[ \frac{1}{G_{-}(Y_s)} + G_{-}(Y_s)^2 \right]
\quad \textrm{with} \quad Y_s=\frac{m^2}{3}\left( \frac{16 \sigma}{a Q^2} \right)^{2/3}, \\
\label{tho5}
M^{(s)}_u&=&3 \left( \sqrt{a} \frac{\sigma}{2} Q \right)^{2/3}.
\end{eqnarray}
Using these formulas, it is easy to check the supplementary relations
(\ref{eq:dual2N4})-(\ref{eq:dual2N6}), (\ref{eq:dual2Nu}), (\ref{eq:dual2Nuu}),
(\ref{eq:dual2Nv}), (\ref{eq:dual2Nuub}).

To check the duality relations between ultrarelativistic and nonrelativistic systems with an example, 
we can consider the following Hamiltonian
\begin{equation}
\label{eq:thurnr}
H = \sum_{i=1}^N \sqrt{\bm p_i^2+m^2} + \textrm{sgn}(\lambda) a \sum_{i=1}^N |\bm r_i - \bm R|^\lambda +
    \textrm{sgn}(\lambda) b \sum_{i<j = 1}^N |\bm r_i - \bm r_j |^\lambda. 
\end{equation}
Defining the quantities
\begin{equation}
\label{eq:AlBl}
A_\lambda = a |\lambda| \left( \frac{N}{Q} \right)^\frac{2-\lambda}{2} 
\quad \textrm{and} \quad
B_\lambda = b |\lambda| N \left( \frac{N-1}{2 Q} \right)^\frac{2-\lambda}{2},
\end{equation}
the AFM solutions, for the nonrelativistic and the ultrarelativistic cases respectively,
are given by \cite{silv10}
\begin{eqnarray}
\label{tho6}
E^{(N)}&=& \frac{\lambda+2}{2 \lambda} Q \left[ \frac{(A_\lambda + B_\lambda)^2}{m^\lambda}
\right]^\frac{1}{\lambda+2}, \\
\label{tho7}
M^{(N)}_u&=&\frac{\lambda+1}{\lambda} \left[ Q^{\lambda+2} N^\lambda (A_\lambda + 
B_\lambda)^2 \right]^\frac{1}{2(\lambda+1)}.
\end{eqnarray}
Using these relations with potentials $W(x)\propto x^\lambda$ and $U(x)\propto x^{2\lambda}$,
the two theorems presented in Sect.~\ref{sec:passnru} can be checked. 

\section*{Acknowledgments}
C. S. would thank the F.R.S.-FNRS for the financial support.


\begin{thebibliography}{99}
\bibitem{dir66} P. A. M. Dirac, \emph{Lectures on Quantum Mechanics} (Belter
Graduate School of Sciences, Yeshiva University, New York, 1966).
\bibitem{brink77} L. Brink, P. Di Vecchia, P. S. Howe, Nucl. Phys. \textbf{B118}, 76 (1977).
\bibitem{deser76} S. Deser, B. Zumino, Phys. Lett. B \textbf{65}, 369 (1976).
\bibitem{polya81} A. M. Polyakov, Phys. Lett. B \textbf{103}, 207 (1981).
\bibitem{bsb08a} B. Silvestre-Brac, C. Semay, F. Buisseret, J. Phys. A
\textbf{41}, 275301 (2008) [arXiv:0802.3601]. 
\bibitem{bsb08b} B. Silvestre-Brac, C. Semay, F. Buisseret, J. Phys. A
\textbf{41}, 425301 (2008) [arXiv:0806.2020]. 
\bibitem{AFMeigen} C. Semay, B. Silvestre-Brac, 
J. Phys. A {\bf 43}, 265302 (2010) [arXiv:1001.1706]. 
\bibitem{bsb09a} B. Silvestre-Brac, C. Semay, F. Buisseret, J. Phys. A \textbf{42},
245301 (2009) [arXiv:0811.0287]. 
\bibitem{Sem09a}  C. Semay, F. Buisseret, B. Silvestre-Brac, Phys. Rev. D
\textbf{79}, 094020 (2009) [arXiv:0812.3291]; 
[arXiv:0901.4614]. 
\bibitem{bsb09c} B. Silvestre-Brac, C. Semay, F. Buisseret, Int. J. Mod. Phys.
A \textbf{24}, 4695 (2009) [arXiv:0903.3181]. 
\bibitem{silv10} B. Silvestre-Brac, C. Semay, F. Buisseret, F. Brau, 
J. Math. Phys. \textbf{51}, 032104 (2010) [arXiv:0908.2829]. 
\bibitem{bsb10b} B. Silvestre-Brac, C. Semay, F. Buisseret, 
arXiv:1101.5222. 
\bibitem{qcd2} G. S. Bali, Phys. Rep. \textbf{343}, 1 (2001) [hep-ph/0001312].
\bibitem{barlnc} F. Buisseret, C. Semay, Phys. Rev. D \textbf{82}, 056008 (2010) [arXiv:1006.4729].
\bibitem{sema01} C. Semay, D. Baye, M. Hesse, B. Silvestre-Brac, Phys. Rev. E \textbf{64}, 016703 (2001). 
\bibitem{hobasis} S. Fleck, B. Silvestre-Brac, J.-M. Richard, Phys. Rev. D
\textbf{38}, 1519 (1988); B. Silvestre-Brac, Few-Body Syst. \textbf{20}, 1 (1996).
\bibitem{brau00} F. Brau, Phys. Rev. D \textbf{62}, 014005 (2000) [hep-ph/0412170].

\end{thebibliography}
\end{document}